\documentclass[a4paper,12pt]{article}
\usepackage[english]{babel}
\usepackage{graphicx,color}
\usepackage{amsmath,amsthm,amssymb,amsfonts}
\usepackage{epsfig}
\usepackage[T1]{fontenc}
\usepackage[cp1250]{inputenc}
\usepackage[bookmarks,colorlinks]{hyperref}
\usepackage{fancyhdr}
\fancypagestyle{plain}{%
\fancyhead{}

}

\hypersetup{colorlinks,%
linkcolor=magenta,%
urlcolor=black,%
citecolor=blue,%
filecolor=green}

\title{On some consequences of the Snyder--Sidharth deformation of Special Relativity}

\author{
\textsl{\L ukasz Andrzej GLINKA\footnote{Electronic address: \href{mailto:laglinka@gmail.com}{\tt{laglinka@gmail.com}}}}
\vspace*{20pt}\\
\emph{Dipartimento di Matematica e Informatica,}\\
\emph{Universit\`{a} degli Studi di Udine,}\\
\emph{Via delle Scienze 206, 33100 Udine, Italia}
}
\date{\today}

\newcommand{\sgn}{\mathrm{sgn}}

\begin{document}

\maketitle
\begin{abstract}
The hypothesis on a minimal scale existence in the Universe leads to noncommutative geometry of Spacetime and thence to a modification of the Special Relativity constraint. Sidharth has deduced that this is equivalent to the Lorentz symmetry violation. This latter consideration was also used by Glashow, Coleman and other scholars though based on purely phenomenological models that have been suggested by the observation of Ultra High Energy Cosmic Rays and Gamma Bursts. On the other hand a parallel development has been the proposal of a small but nonzero photon mass $m_\gamma>0$ by some authors including Sidharth, such a mass being within experimentally allowable limits. This too leads to a small violation of the Lorentz symmetry observable in principle in very high energy gamma rays, as in fact is claimed. In this paper we study the Snyder--Sidharth Hamiltonian and briefly comment the Dirac--Sidharth Hamiltonian, that is a possible explanation for observable violation of the Lorentz symmetry.
\end{abstract}
\newpage

\section{Introduction}

In Special Relativity, the Einstein Hamiltonian constraint holds\footnote{We use the units system $\hbar=c=1$.}
\begin{equation}\label{ec}
  E^2=m^2+p^2,
\end{equation}
and is in fact a minimal quadratic form in the momentum $p$ of a relativistic particle. For this case the Lorentz symmetry is validate. However, recent observations of Ultra High Energy Cosmic Rays and rays from Gamma Bursts seem to suggest a small violation of the Lorentz symmetry \cite{bgs2008}. Also a number of scholars like Glashow, Coleman, and others have considered schemes which depart from the Einstein theory (\ref{ec}). It must be stressed here that these all schemes are purely \emph{ad hoc}.

From a theoretical physics point of view, the problem seems to have a source in the fact that the equation (\ref{ec}) is not the only possible Lorentz invariant quadratic form in momentum $p$. As the example let us consider the natural possibility of deformation of the energetic constraint (\ref{ec}) by simple adding linear term in $p$
\begin{equation}\label{ec1}
  E^2=m^2+p^2+\beta_ip_i,
\end{equation}
where $\beta^i$ is deformation parameter. The modification (\ref{ec1}) nontrivially deforms invariant hyperboloid in the energy-momentum space. This quadratic form can be led by elementary algebraic manipulation to its canonical form
\begin{equation}\label{ec2}
  E^2+\dfrac{|\beta|^2}{4}=\left(\dfrac{\beta_i}{|\beta|}p_i+\dfrac{|\beta|}{2}\right)^2+m^2,~~|\beta|^2=\beta_i\beta_i
\end{equation}
which can be worked out by three different ways of possible interpretation.
\begin{enumerate}
  \item First case is the following identification
\begin{equation}
  \left\{\begin{array}{l}m^2=\dfrac{|\beta|^2}{4}\\E^2=\left(\dfrac{\beta_i}{|\beta|}p_i+\dfrac{|\beta|}{2}\right)^2\end{array}\right.,
\end{equation}
which leads to the physical mass values
\begin{equation}
  m=\pm\dfrac{|\beta|}{2}.
\end{equation}
The solution of the second equation determines energy as
\begin{equation}
\gamma_0E=\gamma_ip_i+m~~,~~{\gamma_0}^2=1~~,~~\gamma_i=\dfrac{\beta_i}{2m}~~,~~\left[\gamma_i,\gamma_j\right]_{+}=2\delta_{ij}.
\end{equation}
This is the \emph{linear} Dirac constraint with the classical Clifford $\gamma$-algebra. After change algebra on four-dimensional
\begin{equation}\label{q1}
\gamma_i\rightarrow\gamma_\mu=\left\{(-\gamma_0,\gamma_i):\left[\gamma_\mu,\gamma_\nu\right]_{+}=-2g_{\mu\nu}\right\},
\end{equation}
and using of the canonical relativistic quantization
\begin{equation}\label{q2}
  (E,p)\rightarrow i\partial_{\mu}=i(-\partial_0,\partial_i),
\end{equation}
the Lorentz symmetry is fully valid for this case.
\item The second possible identification is
\begin{equation}
  \left\{\begin{array}{l}m^2=E^2\\\dfrac{|\beta|^2}{4}=\left(\dfrac{\beta_i}{|\beta|}p_i+\dfrac{|\beta|}{2}\right)^2\end{array}\right..
\end{equation}
In this situation we have the Einstein-type relation between energy and rest mass
\begin{equation}
  E=m,
\end{equation}
where negative mass was rejected as nonphysical, as well as the following relation is established
\begin{equation}
  \dfrac{\beta_i}{|\beta|}p_i=\left\{0,-|\beta|\right\} \longrightarrow p_i=\left\{0,-\beta_i\right\}
\end{equation}
It physically means either rest frame or motion under constant momentum, that is generally an inertial frame. For this case the Lorentz symmetry also holds.
\item The third possibility is
\begin{equation}
  \left\{\begin{array}{l}-E^2=\dfrac{|\beta|^2}{4}\\-m^2=\left(\dfrac{\beta_i}{|\beta|}p_i+\dfrac{|\beta|}{2}\right)^2\end{array}\right..
\end{equation}
which leads to the energy values
\begin{equation}
  \pm i\gamma_0E=\dfrac{|\beta|}{2}~~,~~{\gamma_0}^2=1.
\end{equation}
The solution of the second equation again establishes the Dirac--Clifford classical constraint
\begin{equation}
\gamma_0E=\gamma_ip_i+m~~,~~\gamma_i=-\dfrac{\beta_i}{2E}~~,~~\left\{\gamma_i,\gamma_j\right\}=2\delta_{ij},
\end{equation}
for which the Lorentz symmetry is valid after application of the relativistic quantization procedure (\ref{q1})-(\ref{q2}).
\end{enumerate}

However, the linear deformation (\ref{ec1}) is not the only one which gives the Hamiltonian constraint that is a quadratic form in momentum. It be more general Lorentz invariant constraint is naturally
\begin{equation}\label{gc}
  E^2=m^2+p^2+\left(\beta_i +\delta_i p^2\right)p_i+\alpha p^4,
\end{equation}
where $(\alpha, \beta_i, \delta_i)$ is heptavalent family of deformation parameters. The modification (\ref{gc}) is a good candidate for new type deductions in context of so called ''New Physics''. Especially so the mysterious and almost mystic question as violation of the Lorentz symmetry in high energy physics and cosmology can be studied by \emph{ad hoc} application of the deformations of the Einstein energetic constraint similar to  (\ref{gc}).

\section{The Snyder--Sidharth deformation}

Let us consider now the following deformation of the Einstein constraint (\ref{ec})
\begin{equation}\label{ss}
  E^2=m^2+p^2+\alpha\ell^2p^4,
\end{equation}
where $\ell$ is any minimal physical scale, deduced by Sidharth (Refs. \cite{a,b,c}) in the astroparticle physics context, and investigated by Snyder \cite{snyder} in context of the infrared catastrophe of soft photons in the Compton scattering. In fact this modification follows from the manipulation in phase space of any special relativistic particle
\begin{equation}\label{nd}
  i[p,x]=1+\alpha\ell^2p^2\quad,\quad[x,y]=O(\ell^2)\quad,\quad\alpha\sim1,
\end{equation}
so that we have to deal with the structure of a nondifferentiable manifold, or lattice type model. It must be emphasized that (\ref{nd}) is Lorentz invariant deformation. Sidharth proposed taking into account the Hamiltonian constraint (\ref{ss}), and studying this deformation in wider sense as some type of perturbational series in the minimum scale $\ell$, that can be the Planck scale (or the Compton scale).

As in the case of the linear deformation, the Hamiltonian constraint (\ref{ss}) can be seen easily lead to the canonical form
\begin{equation}\label{qfss}
  E^2+\dfrac{1}{4\alpha\ell^2}=\alpha\ell^2\left(p^2+\dfrac{1}{2\alpha\ell^2}\right)^2+m^2,
\end{equation}
and as previously there are three possible mathematical interpretations of this equation. However, in the considered deformation (\ref{ss}) we have not linear or 3rd-order terms, there are only powers of $p^2$. According to standard rules of Quantum Theory \cite{greiner,peskin} this means that in considered situation the Dirac--Clifford algebraic structure must be absent or hidden.
\begin{enumerate}
\item First, we can interpret the constraint equation (\ref{qfss}) as system of two equations
\begin{equation}
  \left\{\begin{array}{l}m^2=\dfrac{1}{4\alpha\ell^2}\\E^2=\alpha\ell^2\left(p^2+\dfrac{1}{2\alpha\ell^2}\right)^2\end{array}\right.
\end{equation}
The first equation leads to solution that looks like formally as the bosonic string tension
\begin{equation}\label{s1}
  m=\dfrac{1}{2\sqrt{{\alpha}}\ell}.
\end{equation}
Expressing $\alpha\ell^2$ by $m$, one can write the solution of the second equality as follows
\begin{equation}\label{s2}
E=m+\dfrac{p^2}{2m}.
\end{equation}
This is the Hamiltonian of a free point particle in semi-classical mechanics; it is a sum of the Newtonian kinetic energy and the Einstein--Poincare rest energy correction. Interestingly (\ref{s1}) and (\ref{s2}) are consistent if $m$ is the Planck mass, and $\ell$ is the Planck length. It must be remembered that the Planck mass defines a scale where the classical and the quantum meet. For the Planck mass, as is well known the Schwarzschild radius equals to Compton length \cite{kiefer}. In comparison with the non deformed case we have not here higher relativistic corrections. After canonical quantization this is exactly the Schr\"odinger Hamiltonian of free quarks in Quantum Chromo-Dynamics (QCD) because already the quarks are heavy and so non-relativistic \cite{lee}. For the case of vanishing scale and nonzero $\alpha$ as well as for vanishing $\alpha$ and fixed nonzero scale $\ell$, formally $m\equiv\infty$ and energy is also infinite, so that this is a nonphysical black-hole type singularity. For the large scale limit and nonvanishing $\alpha$, there mass spectrum is $m=0$, and for nonzero momentum, this system has infinite energy, so this too is a nonphysical situation. In any case this shows that (\ref{s2}) is compatible with (\ref{nd}).

\item The second case changes the role of energy and mass
\begin{equation}
  \left\{\begin{array}{l}-m^2=\alpha\ell^2\left(p^2+\dfrac{1}{2\alpha\ell^2}\right)^2\\-E^2=\dfrac{1}{4\alpha\ell^2}\end{array}\right.,
\end{equation}
and gives discrete energy spectrum for fixed scale $\ell$. However, it should be mentioned that while (\ref{ss}) with $\alpha>0$ is true for fermions, as was shown by Sidharth \cite{bgs2008}, $\alpha<0$ for bosons. So, for the case of fermions we have here
\begin{equation}\label{ene}
iE=\dfrac{1}{2\sqrt{{\alpha}}\ell},
\end{equation}
as well as the mass one
\begin{equation}\label{m1}
im=\sqrt{{\alpha}}\ell p^2+\dfrac{1}{2\sqrt{{\alpha}}\ell}.
\end{equation}
(rejecting the negative value from). However, one can eliminate scale by energy with the result
\begin{equation}
  m=-\dfrac{p^2}{2E}+E.
\end{equation}
The last equation can be rewritten in the form
\begin{equation}
E^2=mE+\dfrac{p^2}{2},
\end{equation}
and by using of the deformed constraint (\ref{ss}) it yields
\begin{equation}
 m^2+p^2+\alpha\ell^2p^4=mE+\dfrac{p^2}{2}.
\end{equation}
One can find now the energy (not square of energy!), that is $4th$-power in momentum
\begin{equation}\label{pauli}
  E=m+\dfrac{p^2}{2m}+\dfrac{\alpha\ell^2}{m}p^4.
\end{equation}
So, again one can apply the canonical form of a quadratic form
\begin{equation}
  E+\dfrac{1}{16m\alpha\ell^2}=\dfrac{\alpha\ell^2}{m}\left(p^2+\dfrac{1}{4\alpha\ell^2}\right)^2+m,
\end{equation}
and consider three possible cases of identification mass-energy.
\begin{enumerate}
  \item The first obvious interpretation yields
\begin{equation}
  \left\{\begin{array}{l}m=\dfrac{1}{16m\alpha\ell^2}\vspace*{5pt}\\E=\dfrac{\alpha\ell^2}{m}\left(p^2+\dfrac{1}{4\alpha\ell^2}\right)^2\end{array}\right.
\end{equation}
and again solution of the first equation is easy to extract
\begin{equation}\label{mm}
  m=\dfrac{1}{4\sqrt{\alpha}\ell},
\end{equation}
and solution of the second equality can be written in the form of the Pauli Hamiltonian constraint
\begin{equation}
  E=m+\dfrac{p^2}{2m}+\dfrac{p^4}{16m^3}=4\alpha^{3/2}\ell^3\left(p^2+\dfrac{1}{4\alpha\ell^2}\right)^2.
\end{equation}
However, one can easily see by the relation (\ref{ene}) that for this case holds
\begin{equation}
  iE=2m\longrightarrow m^2=-\dfrac{E^2}{2}<0,
\end{equation}
and in consequence values of momentum are non hermitian, so that we have to deal with tachyon.

Moreover, by direct using of the relation (\ref{mm}) together with the formula (\ref{m1}) one establishes the spectrum of momentum $p$ in dependence on the minimal scale $\ell$
\begin{equation}
  p=\mp\dfrac{1}{2\sqrt{\alpha}\ell}\left(\sqrt{\strut{\dfrac{1}{2}\sqrt{\strut{5}}-1}}-i\sqrt{\strut{\dfrac{1}{2}\sqrt{\strut{5}}+1}}\right).
\end{equation}

\item The second case is
  \begin{equation}
  \left\{\begin{array}{l}m=E\\\dfrac{\alpha\ell^2}{m}\left(p^2+\dfrac{1}{4\alpha\ell^2}\right)^2=\dfrac{1}{16m\alpha\ell^2}\end{array}\right.
\end{equation}
Again, solution of the first equation is elementary
\begin{equation}
  E=m=-i\dfrac{1}{2\sqrt{{\alpha}}\ell}\quad,\quad m^2<0,
\end{equation}
and again the tachyon is obtained -- there are particles with momentum spectrum
\begin{equation}
  p=\left\{0,\dfrac{1}{\sqrt{{2\alpha}}\ell}\right\}.
\end{equation}
This is discrete momenta spectrum for fixed scale $\ell$. For running scale this is non compact spectrum, but compactification to the point is done for large scale
\begin{equation}
  \lim_{\ell\rightarrow\infty} p = 0,
\end{equation}
and it is the rest. For $\alpha=0$ and fixed scale $\ell$ there are two singular values of the momentum $p$. For all $\ell\neq0$ and $\alpha\neq0$, the case of nonzero $p$ is related to the existence of tachyon.
\item The third interpretation yields
\begin{equation}
  \left\{\begin{array}{l}-E=\dfrac{1}{16m\alpha\ell^2}\vspace*{5pt}\\-m=\dfrac{\alpha\ell^2}{m}\left(p^2+\dfrac{1}{4\alpha\ell^2}\right)^2\end{array}\right..
\end{equation}
Again by using of the relation (\ref{ene}) we obtain from the first equation
\begin{equation}
  im=\dfrac{1}{8\sqrt{\alpha}\ell},
\end{equation}
and in consequence momentum spectrum is
\begin{equation}
  p=\left\{\pm i\sqrt{\strut{\dfrac{1}{8\alpha}}}\dfrac{1}{\ell},\pm i\sqrt{\strut{\dfrac{3}{8\alpha}}}\dfrac{1}{\ell}\right\},
\end{equation}
so that it is again tachyonic case. By this reason this case - that is the Pauli Hamiltonian constraint with mass related to minimum scale describes tachyon, the hypothetical particle with velocity faster then light.
\end{enumerate}

Results for bosons can be obtained by a simple change
 \begin{equation}\label{cha}
   \alpha\longrightarrow-|\alpha|,
 \end{equation}
 and are more realistic from a current experimental particle physics point of view. In this case the Pauli energetic constraint (\ref{pauli}) has the following form
 \begin{equation}
   E=m+\dfrac{p^2}{2m}-\dfrac{|\alpha|\ell^2}{m}p^4,
 \end{equation}
 and tachyon-like states are absent. We have here the solutions
 \begin{equation}
   E=\dfrac{1}{2\sqrt{|\alpha|}\ell}\quad,\quad m=\sqrt{|\alpha|}\ell p^2+\dfrac{1}{2\sqrt{|\alpha|}\ell},
 \end{equation}
 and three possible situations, that can be easily deduced from the fermionic case presented above, with the change (\ref{cha}).

 \item The third possible solution of the Snyder--Sidharth Hamiltonian constraint can be constructed by the system of equations
 \begin{equation}
  \left\{\begin{array}{l}E^2=m^2\vspace*{5pt}\\\dfrac{1}{4\alpha\ell^2}=\alpha\ell^2\left(p^2+\dfrac{1}{2\alpha\ell^2}\right)^2\end{array}\right..
\end{equation}
First equation gives standard Einstein relation
\begin{equation}
  E=m,
\end{equation}
and the second equality leads to the discrete momentum spectrum
\begin{equation}
  p=\left\{0,\dfrac{1}{\sqrt{{\alpha}}}\dfrac{1}{\ell}\right\}.
\end{equation}
For fixed scale $\ell$ this is discrete spectrum. For running scale this is non compact spectrum, but compactification to the point spectrum is done in the large scale limit
\begin{equation}
  \lim_{\ell\rightarrow\infty} p = 0,
\end{equation}
and it is the rest. For $\alpha=0$ and fixed scale $\ell$ there are two singular values of the momentum $p$. For all $\ell\neq0$ and $\alpha\neq0$, the case of nonzero $p$ is related to the existence of a relativistic particle.
\end{enumerate}

\section{Scale-modified Compton effect}
Let us consider now the case of the Compton effect with the Sidharth Hamiltonian constraint (\ref{ss}). In the standard case, in the CM system, wave vector of outgoing photon $k$ is related to wave vector of incoming photon $k_0$ scattered on the electron with mass $m$ by the relation
\begin{equation}
k=\dfrac{mk_0}{m+k_0(1-\cos\theta)}.
\end{equation}
The point is the modification of this wave vector according to the idea
\begin{equation}
  \omega^2_{eff}=m^2+k^2_{eff},
\end{equation}
where $k_{eff}$ is the corrected wave vector
\begin{equation}
  k^2_{eff}=k'^2+\alpha\ell^2k'^4\quad,\quad\alpha=-|\alpha|.
\end{equation}
Effectively one can obtain the relation
\begin{equation}
  k'=k+\epsilon,
\end{equation}
where $\epsilon$ is the correction from non vanishing scale $\ell$
\begin{equation}\label{37}
  \epsilon=[Q^2+2mQ]^2\dfrac{\omega}{\omega_0}\dfrac{\alpha\ell^2}{2m},
\end{equation}
$Q$ is the difference
\begin{equation}\label{38}
  Q=k-k_0,
\end{equation}
and the frequencies were introduced
\begin{equation}
  \omega_0=\dfrac{k_0}{m}\quad,\quad\omega=\dfrac{k}{m}=\dfrac{\omega_0}{1+\omega_0(1-\cos\theta)}.
\end{equation}
According to energy conservation law one can establish the mass of the photon as
\begin{equation}
  m_\gamma=k_{eff}-k'=k'\left(\sqrt{\strut{1+\alpha\ell^2k'^2}}-1\right),
\end{equation}
and it is non zeroth for non vanishing scales $\ell\neq0$. This relation leads to the formula
\begin{equation}
  m_\gamma+k'=k'\sqrt{\strut{1+\alpha\ell^2k'^2}},
\end{equation}
which can be rewritten in the form of equation for $k'$
\begin{equation}
  \alpha\ell^2k'^4-2m_\gamma k'-m_\gamma^2=0.
\end{equation}
For finite photon mass $m_\gamma>0$ and nonzeroth scale $\ell\neq0$ this equation has complex solution
\begin{equation}
\omega'=\dfrac{k'}{m}=\omega_R+i\omega_I,
\end{equation}
where $\omega_R$ and $\omega_I$ are real and imaginary parts of $\omega'$
\begin{eqnarray}
 \omega_R&=&\pm\dfrac{m_\gamma}{m}\dfrac{1}{\eta\xi(\eta)},\\
 \omega_I&=&\mp\omega_R\sqrt{\strut{1+\dfrac{1}{3}\xi^3(\eta)}}.
\end{eqnarray}
Here we have introduced the function $\xi(\eta)$
\begin{equation}
  \xi(\eta)=\sqrt{\strut{\dfrac{18^{1/3}\left(\strut{1+\sgn(\alpha)\sqrt{\strut{1+\dfrac{4}{9}\eta^3}}}\right)^{1/3}}{\left(1+\sgn(\alpha)\sqrt{1+\dfrac{4}{9}\eta^3}\right)^{2/3}-12^{1/3}\eta}}},
\end{equation}
where $\eta$ is the parameter
\begin{equation}
  \eta^3=\dfrac{4}{3}\alpha\ell^2m_\gamma^2.
\end{equation}
One can easily establish $\epsilon$ by difference of frequencies as
\begin{equation}
  \dfrac{\epsilon(\ell)}{m}=\omega'-\omega,
\end{equation}
and by similarly one can determine the constant $Q$ directly from (\ref{37})
\begin{equation}
Q(\ell)=-m\left(1\mp\sqrt{\strut{1+\dfrac{2}{3}\dfrac{m_\gamma}{m}\sqrt{\strut{\dfrac{6\omega_0}{\eta^3}}}\sqrt{\strut{\dfrac{\epsilon(\ell)/m}{\omega}}}}}\right).
\end{equation}
By using of the relation (\ref{38}) rewritten in terms of the frequencies
\begin{equation}
  \dfrac{Q(\ell)}{m}=\omega-\omega_0,
\end{equation}
one can finally establish the energy gap $\epsilon$ as
\begin{equation}
  \dfrac{\epsilon(\ell)}{m}=\omega'-\omega_0+1\mp\sqrt{\strut{1+\dfrac{2}{3}\sqrt{\strut{\dfrac{6}{\eta^3}}}\sqrt{\strut{\dfrac{\omega_0}{\omega}}}\dfrac{m_\gamma}{m}\sqrt{\strut{\omega'-\omega}}}}.
\end{equation}
Since $\omega'$ is a complex number, one can write the gap energy by employing of the polar representation
\begin{equation}\label{me}
  \dfrac{\epsilon(\ell)}{m}=\sqrt{\strut{\epsilon_R^2(\ell)+\epsilon_I^2(\ell)}}\exp\left(i\vartheta(\ell)\right)\quad,\quad\vartheta(\ell)\equiv\mathrm{arg}\epsilon(\ell)
\end{equation}
where $\vartheta$ is a phase
\begin{equation}\label{ph}
  \vartheta=n\cdot\arctan\dfrac{\omega_I\pm\sqrt{\strut{-\dfrac{1+ax}{2}+\dfrac{1}{2}\sqrt{\strut{(1+ax)^2+y^2}}}}}{\omega_R-\omega_0+1\pm\sqrt{\strut{\dfrac{1+ax}{2}+\dfrac{1}{2}\sqrt{\strut{(1+ax)^2+y^2}}}}},
\end{equation}
where $n$ is any integer, $\epsilon_R$ is real part of the gap energy
\begin{equation}
  \epsilon_R=\omega_R-\omega_0+1\pm\sqrt{\strut{\dfrac{1+ax}{2}+\dfrac{1}{2}\sqrt{\strut{(1+ax)^2+y^2}}}},
\end{equation}
and $\epsilon_I$ is its imaginary part
\begin{equation}
  \epsilon_I=\omega_I\pm\sqrt{\strut{-\dfrac{1+ax}{2}+\dfrac{1}{2}\sqrt{\strut{(1+ax)^2+y^2}}}}.
\end{equation}
Here, for shorten notation, we have introduced the abbreviations
\begin{equation}
  a\equiv \dfrac{2}{3}\sqrt{\strut{\dfrac{6}{\eta^3}}}\sqrt{\strut{\dfrac{\omega_0}{\omega}}}\dfrac{m_\gamma}{m},
\end{equation}
\begin{eqnarray}
  x&\equiv&\sqrt{\strut{\dfrac{\omega_R-\omega}{2}+\dfrac{1}{2}\sqrt{\strut{\left(\omega_R-\omega\right)^2+\omega_I^2}}}},\\
  y&\equiv&\sqrt{\strut{-\dfrac{\omega_R-\omega}{2}+\dfrac{1}{2}\sqrt{\strut{\left(\omega_R-\omega\right)^2+\omega_I^2}}}},
\end{eqnarray}
which are consistent for $\omega_R\geq\omega$, $\omega_I\geq0$, $1+ax\geq0$. The existence of the scale-dependent nonzero phase (\ref{ph}) reflects the property of multiply connected space.

However the fundamental part of multi-energy (\ref{me}) can be established by the Bohr--Sommerfeld quantization rule for the phase of energy, \emph{i.e.}
\begin{equation}
  \vartheta=n'\cdot2\pi,
\end{equation}
where $n'$ is any integer. In this case we have simply $\epsilon_I=0$, and the total energy gap is determined as
\begin{equation}
  \dfrac{\epsilon(\ell)}{m}=\left|\epsilon_R(\ell)\right|=\left|\omega_R-\omega_0+1\pm\sqrt{\strut{\omega_I^2+ax+1}}\right|.
\end{equation}
Taking the minimal scale as the Planck scale $\ell=\ell_P=\sqrt{\strut{\dfrac{\hslash G}{c^3}}}\approx 1.61625 \times 10^{-35} m$ (in SI units) one receives the fundamental energy gap as $1$ eV. The multiple value of an energy gap can be interpreted as the multiple connected property of Spacetime \cite{a}.
\section{Conclusion}

The supposition about non vanishing photon mass was deduced based on a background Dark Energy or the Zero Point Field (Cf. ref. \cite{a}) at the Planck scale. On the other hand employing the Planck scale as the minimum scale, it is known that spacetime geometry becomes nondifferentiable and noncommutative. In fact it modifies the usual Lorentz symmetry existing in the Klein--Gordon and Dirac equations.

The photon mass within the experimental constraints, but also leads to observable results in the High Energy Gamma Ray spectrum and Gamma Bursts astrophysics. We are full of hopes that NASA's GLAST satellite will throw further light on this. We emphasize that the energy--momentum relation (\ref{ss}) leads to a Dirac type Hamiltonian, the Dirac--Sidharth Hamiltonian with interesting consequences in the Ultra High Energy regime \cite{a,b}.

Formally, we have shown that Special Relativity modified by the noncommutative geometry of Spacetime (\ref{ss}) can be resolved (\emph{i.e.} Hamiltonian can be established within) by some nontrivial classical ways. It means that the Snyder--Sidharth deformation of the Einstein theory leads to some nontrivial quantum theories, that are dependent on relations of energy $E$, spatial momentum $p$ and mass $m$ of a relativistic particle and the minimal scale $\ell$.

\section*{Acknowledgements}
The author thanks A. De Angelis and F. Freschi for full hospitality at the University of Udine.

\end{document}